\documentclass[a4paper, twocolumn]{article}

\usepackage{algorithm}
\usepackage{algorithmic}
\usepackage{booktabs} 
\usepackage{hyperref}
\usepackage{graphicx}
\usepackage{authblk}

\title{\textbf{MPSUM: Entity Summarization with Predicate-based Matching}}

\author[1,2]{Dongjun Wei}
\author[1,2]{Shiyuan Gao}
\author[1,2]{Yaxin Liu}
\author[1,2]{Zhibing Liu}
\author[1]{Longtao Hang\thanks{Correspondence to Longtao Huang.}}
\affil[1]{Institute of Information Engineering, Chinese Academy of Sciences, China}
\affil[2]{School of Cyber Security, University of Chinese Academy of Sciences, China}
\affil[ ]{\{\url{weidongjun, gaoshiyuan, liuyaxin, liuzhibing, huanglongtao}\}\url{@iie.ac.cn}}

\date{}

\begin{document}

\maketitle

    \begin{abstract}
        With the development of Semantic Web, entity summarization has become an emerging task to
        generate concrete summaries for real world entities. To solve this problem, we propose an
        approach named MPSUM that extends a probabilistic topic model by integrating the idea of
        predicate-uniqueness and object-importance for ranking triples. The approach aims atwebview-panel:webview-panel/webview-0fb291c9-310d-42e2-a420-ed1a22b0a5ac
        generating brief but representative summaries for entities. We compare our approach with
        the state-of-the-art methods using DBpedia and LinkedMDB datasets. The experimental
        results show that our work improves the quality of entity summarization.
        The source code and outputs are available at \url{https://github.com/WeiDongjunGabriel/MPSUM}\footnote{This paper was accepted in EYRE@CIKM'2018.}.
    \end{abstract}
	
    \section{Introduction}

    Linked Open Data (LOD) can describe entities on the Semantic Web using
    Uniform Resource Identifiers (URIs) or Resource Description Framework
    (RDF). Therefore, LOD is regarded as a collection of entity
    descriptions and has formed many public datasets, such as
    DBpedia\cite{bizer2009dbpedia} and LinkedMDB\cite{consens2008managing}.
    An RDF triple is in the form of $<subject, predicate,
    object>$. However, lengthy descriptions will take much time for users to
    comprehend and identify the underlying entities. To solve this problem,
    entity summarization has been proposed to generate a set of
    descriptions that are brief but effective to acquire enough information
    for quick comprehension. In this paper, we propose a method called MPSUM
    based on LDA model to identify the $top-k$ representative triples as
    summaries for entities. Apart from ranking triples based on their
    probability distributions, we propose a novel method for triples ranking with consideration of the
    importance of objects and uniqueness of predicates in RDF data.
	
    \section{Related Work}

    RELIN\cite{cheng2011relin} is a variant of the random surfer model for
    ranking features mainly based on relatedness and informativeness for
    quick identification of entities. DIVERSUM\cite{sydow2010diversum}
    solves the problem of diversified entity summarization in RDF-like
    knowledge graphs by incorporating the notion of diversification into
    the summarizing algorithm. FACES-E\cite{gunaratna2016gleaning} is able
    to match a suitable class from existing ontology classes
    set, which extends FACES to generate entity summaries in the way of
    gleaning and ranking object and datatype properties. CD\cite{xu2016cd}
    formulates entity summarization as a binary quadratic knapsack problem
    to solve. FACES\cite{gunaratna2015faces} improves the quality of
    entity summaries by taking the diversification of the relation types
    into consideration, introduces the concept of Cobweb clustering
    algorithm to partition features and rank them.
    LinkSUM\cite{thalhammer2016linksum} partitions the semantic links of
    each entity to rank features and is interfaced via the SUMMA entity
    summarization API. ES-LDA\cite{pouriyeh2017lda} is a probabilistic
    topic model based on LDA to generate representative summaries for
    entities and outperforms LinkSUM and FACES.
	
    \section{Preliminaries}

    \subsection{Resource Description Framework (RDF)}

    RDF is a data modeling language of Semantic Web and is widely used to
    describe entities or resources. An RDF data graph is a set of entities (nodes)
    and relations (edges) between them, which is referred to a collection of
    triples where each triple $t$ consists of a subject $s$, predicate $p$,
    object $o$, in the form of $<s, p, o>$.

    {\bfseries Document}
    \label{df:doc}
    A document $d$ is defined as a collection of triples,
    $d = \{ t_1, t_2, ..., t_n \}$, that describes a single entity $e$.
    Therefore, all triples of a document $d$ have the same subject.

	{\bfseries Entity summarization}
    \label{df:es}
    Given an entity $e$ and a positive integer $k$, a summary of the entity
    $e$ is $ES(e, k)$, is the $top-k$ subset of all predicates and
    corresponding objects that are most relevant to that entity.

    \subsection{Latent Dirichlet Allocation (LDA)}

    The Latent Dirichlet Allocation (LDA)\cite{wei2006lda} is an unsupervised
    machine learning technique that can be used to identify latent topics
    from a collection of documents. It uses a "bag of words" approach that
    regards each document as a word frequency vector, transforming the text
    information into digital information for modeling. LDA generates the words
    in a two-stage process: words are generated from topics and topics are
    generated from documents.
	
    \section{Problem Statement}

    \subsection{Problem Definition}

    An RDF document is in the form of a set of triples consisting of a subject with all predicates and corresponding
    objects related to a specific entity. In this paper, given
    an RDF dataset, entities are described by sets of their properties
    and corresponding objects. Our objective is to select $top-k$
    representative triples to best describe each entity.

    \subsection{Supplementing RDF Data}

    Based on two methods proposed to supplement the RDF data in
    ES-LDA\cite{pouriyeh2017lda}, we enrich the information of each document
    by adding categories of the objects in DBpedia directly to the
    document and expanding each document by increasing the frequency of
    each object with the number of its categories.

    \subsection{Proposed Model}

    MPSUM is an extended model of LDA. The key idea behind our model is
    two-fold: (1) we apply the concept of LDA topic modeling into entity
    summarization; and (2) a novel method is proposed to rank triples and
    improve the performance of LDA in entity summarization.

\begin{figure}
      \begin{center}
	    \includegraphics[width=5cm]{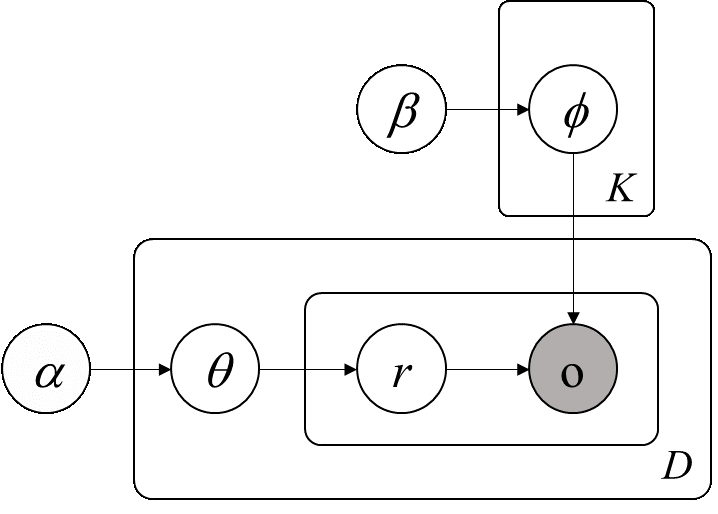}
	  \end{center}
      \caption{Entity Summarization Using LDA}
      \label{fig:ESM}
    \end{figure}

    The number of topics in MPSUM $K$ is set to be the number of
    unique predicates in the corpus. Let $D=\{d_1, d_2, ..., d_{\left|D\right|}\}$
    be a corpus of documents. In our model, each document $d$ is a multinomial
    distribution over the predicate $r$ and each predicate is a multinomial
    distribution over the object $o$ from the Dirichlet prior $\alpha$ and $\beta$
    respectively, in addition, $\theta$ and $\phi$ are latent variables. The plate
    notation is shown in Figure \ref{fig:ESM}.

    ES-LDA\cite{pouriyeh2017lda} only uses probability distributions when ranks
    RDF triples. However, Bian et al.\cite{bian2014research} combined
    topic-importance and topic-distribution for sentence-ranking problem and got a better
    performance. Since each element of an RDF triple has different role, the
    influence of various elements in RDF triples cannot be regarded as the same.
    Each document can be deemed as "a bag of objects" because it owns the same
    subjects as mentioned above in {\emph Definition \ref{df:doc}}. Therefore, a method
    called MP (match up objects and RDF triples based on predicates) has been
    proposed to rank the RDF triples via taking the predicate-uniqueness and
    object-importance into consideration.

    \begin{algorithm}[!h]
    	\caption{MP}
    	\begin{algorithmic}[1]
    		\STATE Retrieve topic words $words$ and RDF triples $triples$ in document $d$ from trained model
        \STATE Initialize predicates' list $predicates$ in $d$
        \FOR{topic word $tw$ in $words$}
        \IF{$tw$ is in $d$}
        \FOR{RDF triple $rp$ in $triples$}
        \STATE predicate $p \leftarrow extractor(rp)$
        \IF{$tw$ matches $rp$ and $p$ is not in $predicates$}
        \STATE Add $p$ into $predicates$
        \STATE Output $rp$ and remove it from $triples$
        \ENDIF
        \ENDFOR
        \ENDIF
        \ENDFOR
        \FOR{RDF triple $rp$ in $triples$}
        \STATE predicate $p \leftarrow extractor(rp$)
        \IF{$p$ is not in $predicates$}
        \STATE Add $p$ to $predicates$
        \STATE Output $rp$ and remove it from $triples$
        \ENDIF
        \ENDFOR
        \FOR{RDF triple $rp$ in $triples$}
        \STATE Output $rp$ and remove it from $triples$
        \ENDFOR
    	\end{algorithmic}
    \end{algorithm}

    Our model repeats the following process of MP for each document.

    {\bfseries Step 1:} Initialize the predicate set and enumerate all the RDF
    triples of an entity to identify the triples of which the objects
    are ranked based on probabilities. Then extract the corresponding
    predicate and add it into the aforementioned predicate set and output the triple
    when the predicate is not in the collection.

    {\bfseries Step 2:} Select the next triple from the remaining ones until all the
    objects complete the matching work, extract their predicates to compare with
    the predicate set generated from Step 1. If the predicate is not in the
    collection, add it into the current predicate set and output the related triple.

    {\bfseries Step 3:} Output the rest triples in order.

    For example, after applying LDA method for training, top-10 predicates are
    selected according to the probability of each object within the subject which has the
    highest probability and corresponding triples of a given subject, including
    \emph{broadcastArea, broadcastArea, callsignMeaning, programmeFormat,
    type, type, label, name, type} and \emph{homepage}. However, apart from the
    probabilities of objects, MP method takes the information of subjects which
    is based on predicates into consideration, and generates better training
    effects, since the top-10 predicates are \emph{broadcastArea, callsignMeaning,
    programmeFormat, label, name, type, subject, homepage, slogan} and
    \emph{type}. The generative process of MP is shown in {\bfseries Algorithm 1}.

    \subsection{Estimating Posterior Inference}

    Since it's difficult to acquire the posterior inference of the LDA, it needs
    to find an algorithm for estimating the posterior inference. There are many
    existing methods including variational EM\cite{blei2003latent} and Gibbs
    sampling\cite{robert2013monte}. EM\cite{blei2003latent} is a maximum
    likelihood estimation method including probabilistic model parameters of
    latent variables. Gibbs sampling\cite{robert2013monte} is a Markov Chain
    Monte Carlo algorithm, which constructs a Markov chain over the latent
    variables in the model and converges to the posterior distribution, after
    a number of iterations. However, TF-IDF\cite{huang2011text} is a statistical
    method to assess the importance of a word for a file set or one of the
    files in a corpus. In our case, we evaluated our model using EM, Gibbs
    sampling and Gibbs sampling with TF-IDF to estimate posterior inference and
    the results demonstrate that the Gibbs sampling shows the best performance.
	
    \section{Experiments and Results}

    {\bfseries Data Preprocessing:} Excessive work on RDF triples would introduce more
    corresponding triples to the topics that have higher probabilities and reduce
    the precision of MP method. Then we apply a concise extraction
    algorithm for objects to adapt to the algorithm for estimating the posterior
    inference. The extraction algorithm first acquires RDF
    triples which conclude the selected objects, then intercepts the part from
    the last '\#' or '/' of RDF to the end. For example, we
    extract \emph{"broadcaster"} after lowering the capital letters
    from '$http://dbpedia.org/ontology/Broadcaster$'.

    {\bfseries Corpus Enlarging: }As mentioned in section 4.2, we
    supplement the RDF data via adding categories of objects and repeating the
    topics in each document to deal with common RDF data problems
    including \emph{sparseness, lack of context}, etc. Compared with the methods
    of supplementing RDF data in ES-LDA model, the results of our model
    outperforms better.

    {\bfseries Inference Algorithm:} In section 4.4, we have introduced three
    algorithms to estimate the posterior inference including EM,
    Gibbs sampling and Gibbs sampling with TF-IDF. All of the above methods have
    been used in our experiments respectively, and the Gibbs sampling shows the
    best results.

    {\bfseries Model Training:} In the training process, it's essential to set
    proper hyperparameters, $\alpha$ and $\beta$. We conduct experiments and finally find that when $\alpha=(E/20)/R$,
    , we can get a satisfying result. $E$ is the total number of unique entities and$R$ is the total number
    of unique predicates. We test 3 configurations as table \ref{tab:Mdd}.
    As the results show, taking the total number of
    entities into account can optimize training effects. When the value of $\beta$
    is set to $0.01$, DBpedia has the best result since it contains enrich corpus.
    However, for LinkedMDB, its corpus is insufficient, when $\beta = 50 / R$,
    the result is better.

    \begin{table}
      \caption{MAP values on different configurations}
      \label{tab:Mdd}
      \begin{center}
        \begin{tabular}{p{2cm}p{1.5cm}p{1.5cm}}
          \toprule
          dbpedia: & top5 & top10 \\
          \midrule
          config\_1: & 0.389 & 0.463 \\
          config\_2: & {\bfseries 0.396} & {\bfseries 0.568}\\
          config\_3: & 0.379 & 0.554\\
          \bottomrule
          \\
          \toprule
          lmdb: & top5 & top10 \\
          \midrule
          config\_1: & 0.370 & 0.474 \\
          config\_2: & 0.370 & 0.476 \\
          config\_3: & {\bfseries 0.371} & {\bfseries 0.576} \\
          \bottomrule
        \end{tabular}
      \end{center}
    \end{table}

    The DBpedia and LinkedMDB datasets are chosen for our experiments. We
    evaluate the F-measures (the harmonic average of the precision and recall)
    and MAP (Mean Average Precision) to compare our MPSUM model with other
    state-of-the-art approaches including RELIN, DIVERSUM, FACES-E, CD,
    FACES and LinkSUM ,and the results are in Table \ref{tab:Fm} and
    Table \ref{tab:MAP} respectively. From the results, we can observe that MPSUM performs best on all cases except the top-10 in DBpedia.As the results show,RELIN and LinkSUM could not meet the diversity requirement in the
    summarization process. FACES discount literals in entity summarization
    while FACES-E and RELIN take literals into account. Our approach maintains both diversity
    and relevancy, while representing each entity through $top-k$ predicates. Above all, MPSUM outperforms the selected approaches on overall datasets.
    \begin{table*}[!ht]
      \caption{F-measure of selected entity summarizers under their best parameter settings}
      \label{tab:Fm}
      \begin{minipage}{\textwidth}
        \begin{center}
          \begin{tabular}{ccccccc}
            \toprule
            & \multicolumn{2}{c}{DBpedia} & \multicolumn{2}{c}{LinkedMDB} & \multicolumn{2}{c}{All}\\
            & k=5 & k=10 & k=5 & k=10 & k=5 & k=10\\
            \midrule
            RELIN & 0.250$_{\lambda=1.00}$ & 0.468$_{\lambda=1.00}$ & 0.210$_{\lambda=1.00}$ & 0.260$_{\lambda=1.00}$ & 0.239$_{\lambda=1.00}$ & 0.409$_{\lambda=1.00}$\\
            DIVERSUM & 0.260 & 0.522 & 0.222 & 0.365 & 0.239 & 0.477\\
            CD & 0.299$_{\gamma=0.47}$ & {\bfseries 0.531}$_{\gamma=0.23}$ & 0.215$_{\gamma=1.00}$ & 0.326$_{\gamma=1.00}$ & 0.267$_{\gamma=0.52}$ & 0.467$_{\gamma=0.16}$\\
            FACES-E & 0.285 & 0.527 & 0.252 & 0.348 & 0.276 & 0.476\\
            FACES & 0.272 & 0.439 & 0.160 & 0.259 & 0.240 & 0.388\\
            LinkSUM & 0.290$_{\alpha=0.01}$ & 0.498$_{\alpha=0.04}$ & 0.117$_{\alpha=1.00}$ & 0.255$_{\alpha=1.00}$ & 0.240$_{\alpha=0.01}$ & 0.428$_{\alpha=0.04}$\\
            MPSUM & {\bfseries 0.313} & 0.522 & {\bfseries 0.270} & {\bfseries 0.440} & {\bfseries 0.300} & {\bfseries 0.499}\\
            Better\footnote{By how much we are better than the best result of all other methods.} & 0.014 & - & 0.018 & 0.075 & 0.024 & 0.022\\
            \bottomrule
          \end{tabular}
        \end{center}
      \end{minipage}
    \end{table*}

    \begin{table*}[!ht]
      \caption{MAP of selected entity summarizers under their best parameter settings}
      \label{tab:MAP}
      \begin{minipage}{\textwidth}
        \begin{center}
          \begin{tabular}{ccccccc}
            \toprule
            & \multicolumn{2}{c}{DBpedia} & \multicolumn{2}{c}{LinkedMDB} & \multicolumn{2}{c}{All}\\
            & k=5 & k=10 & k=5 & k=10 & k=5 & k=10\\
            \midrule
            RELIN & 0.348$_{\lambda=1.00}$ & 0.532$_{\lambda=1.00}$ & 0.243$_{\lambda=1.00}$ & 0.337$_{\lambda=1.00}$ & 0.318$_{\lambda=1.00}$ & 0.476$_{\lambda=1.00}$\\
            DIVERSUM & 0.316 & 0.511 & 0.269 & 0.388 & 0.302 & 0.476\\
            CD & - & - & - & - & - & -\\
            FACES-E & 0.354 & 0.529 & 0.258 & 0.361 & 0.326 & 0.481\\
            FACES & 0.247 & 0.386 & 0.140 & 0.261 & 0.261 & 0.351\\
            LinkSUM & 0.246$_{\alpha=0.25}$ & 0.386$_{\alpha=0.03}$ & 0.120$_{\alpha=1.00}$ & 0.254$_{\alpha=1.00}$ & 0.210$_{\alpha=0.25}$ & 0.348$_{\alpha=0.03}$\\
            MPSUM & {\bfseries 0.396} & {\bfseries 0.568} & {\bfseries 0.371} & {\bfseries 0.476} & {\bfseries 0.389} & {\bfseries 0.542}\\
            Better\footnote{By how much we are better than the best result of all other methods.} & 0.042 & 0.036 & 0.102 & 0.088 & 0.063 & 0.061\\
            \bottomrule
          \end{tabular}
        \end{center}
      \end{minipage}
    \end{table*}
	
    \section{Conclusion}

    In this paper, we put forward an LDA-based model MPSUM for entity
    summarization. In our method, we propose to increase the
    frequency of words by adding categories of the objects to supplement
    RDF data, which is an improvement of ES-LDA\cite{pouriyeh2017lda}. Besides,
    a novel method called MP has been proposed to rank triples with consideration of the
    importance of objects and uniqueness of predicates in RDF data.
    We utilize three algorithms (EM, Gibbs sampling, Gibbs sampling with
    TF-IDF) to estimate posterior inference and finally take advantage of
    Gibbs sampling for experiments and comparisons. The experimental results
    of our approach for entity summarization are quite promising. It
    performs better than 5 state-of-the-art techniques to generate
    summaries.

    \section{Future Work}
    In this paper, we
    select the topic with the maximum probability. For further work, the
    selection of topics can be implemented by proportion to apply more topics
    for ranking RDF triples. Moreover, other proper methods to enlarge RDF
    data remain to be explored to improve the quality of representative triples
    for entity summarization.

	\section*{Acknowledgements}
	
		This research is supported in part by the National Natural Science Foundation of China under Grant No. 61702500.

    \bibliographystyle{plain} 
    \bibliography{ref}

\begin{thebibliography}{10}

\bibitem{bian2014research}
Jinqiang Bian, Zengru Jiang, and Qian Chen.
\newblock Research on multi-document summarization based on lda topic model.
\newblock In {\em Intelligent Human-Machine Systems and Cybernetics (IHMSC),
  2014 Sixth International Conference on}, volume~2, pages 113--116. IEEE,
  2014.

\bibitem{bizer2009dbpedia}
Christian Bizer, Jens Lehmann, Georgi Kobilarov, S{\"o}ren Auer, Christian
  Becker, Richard Cyganiak, and Sebastian Hellmann.
\newblock Dbpedia-a crystallization point for the web of data.
\newblock {\em Web Semantics: science, services and agents on the world wide
  web}, 7(3):154--165, 2009.

\bibitem{blei2003latent}
David~M Blei, Andrew~Y Ng, and Michael~I Jordan.
\newblock Latent dirichlet allocation.
\newblock {\em Journal of machine Learning research}, 3(Jan):993--1022, 2003.

\bibitem{cheng2011relin}
Gong Cheng, Thanh Tran, and Yuzhong Qu.
\newblock Relin: relatedness and informativeness-based centrality for entity
  summarization.
\newblock In {\em International Semantic Web Conference}, pages 114--129.
  Springer, 2011.

\bibitem{consens2008managing}
Mariano~P Consens.
\newblock Managing linked data on the web: The linkedmdb showcase.
\newblock In {\em Web Conference, 2008. LA-WEB'08., Latin American}, pages
  1--2. IEEE, 2008.

\bibitem{gunaratna2016gleaning}
Kalpa Gunaratna, Krishnaprasad Thirunarayan, Amit Sheth, and Gong Cheng.
\newblock Gleaning types for literals in rdf triples with application to entity
  summarization.
\newblock In {\em International Semantic Web Conference}, pages 85--100.
  Springer, 2016.

\bibitem{gunaratna2015faces}
Kalpa Gunaratna, Krishnaprasad Thirunarayan, and Amit~P Sheth.
\newblock Faces: Diversity-aware entity summarization using incremental
  hierarchical conceptual clustering.
\newblock In {\em AAAI}, pages 116--122, 2015.

\bibitem{huang2011text}
Cheng-Hui Huang, Jian Yin, and Fang Hou.
\newblock A text similarity measurement combining word semantic information
  with tf-idf method.
\newblock {\em Jisuanji Xuebao(Chinese Journal of Computers)}, 34(5):856--864,
  2011.

\bibitem{pouriyeh2017lda}
Seyedamin Pouriyeh, Mehdi Allahyari, Krzysztof Kochut, Gong Cheng, and
  Hamid~Reza Arabnia.
\newblock Es-lda: Entity summarization using knowledge-based topic modeling.
\newblock In {\em Proceedings of the Eighth International Joint Conference on
  Natural Language Processing (Volume 1: Long Papers)}, volume~1, pages
  316--325, 2017.

\bibitem{robert2013monte}
Christian Robert and George Casella.
\newblock {\em Monte Carlo statistical methods}.
\newblock Springer Science \& Business Media, 2013.

\bibitem{sydow2010diversum}
Marcin Sydow, Mariusz Piku{\l}a, and Ralf Schenkel.
\newblock Diversum: Towards diversified summarisation of entities in knowledge
  graphs.
\newblock In {\em Data Engineering Workshops (ICDEW), 2010 IEEE 26th
  International Conference on}, pages 221--226. IEEE, 2010.

\bibitem{thalhammer2016linksum}
Andreas Thalhammer, Nelia Lasierra, and Achim Rettinger.
\newblock Linksum: using link analysis to summarize entity data.
\newblock In {\em International Conference on Web Engineering}, pages 244--261.
  Springer, 2016.

\bibitem{wei2006lda}
Xing Wei and W~Bruce Croft.
\newblock Lda-based document models for ad-hoc retrieval.
\newblock In {\em Proceedings of the 29th annual international ACM SIGIR
  conference on Research and development in information retrieval}, pages
  178--185. ACM, 2006.

\bibitem{xu2016cd}
Danyun Xu, Liang Zheng, and Yuzhong Qu.
\newblock Cd at ensec 2016: Generating characteristic and diverse entity
  summaries.
\newblock In {\em SumPre@ ESWC}, 2016.

\end{thebibliography}

\end{document}